\documentclass[11pt]{article}
\usepackage[utf8]{inputenc}
\usepackage[margin=1in]{geometry}
\usepackage{graphicx}
\usepackage{enumitem}
\usepackage{hyperref}
\usepackage{pythonhighlight}
\usepackage{multirow}
\usepackage{tabularx}
\usepackage{longtable,booktabs}
\usepackage[authoryear]{natbib}
\usepackage[toc,page]{appendix}

\title{SpectroMap: Peak detection algorithm for audio fingerprinting}
\author{
\href{https://orcid.org/0000-0001-8332-0381}{Aar\'{o}n~L\'{o}pez-Garc\'{i}a\hspace{1mm}\includegraphics[scale=0.1]{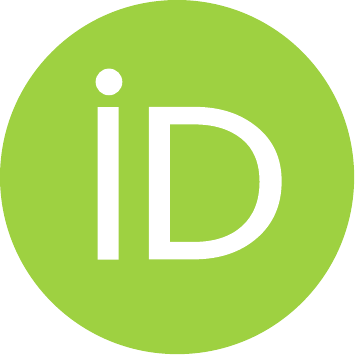}}\\
	Department of Computer Science\\
	University of Valencia\\
	\href{mailto:logara8@alumni.uv.es}{\texttt{logara8@alumni.uv.es}}
}
\date{September 2022}

\begin{document}

\maketitle

\begin{abstract}
\noindent Audio fingerprinting is a technique used to identify and match audio recordings based on their unique characteristics. It involves creating a condensed representation of an audio signal that can be used to quickly compare and match against other audio recordings. The fingerprinting process involves analyzing the audio signal to extract certain features, such as spectral content, tempo, and rhythm, among other things. In this paper, we present SpectroMap, an open-source GitHub repository for audio fingerprinting written in Python programming language. It is composed of a peak search algorithm that extracts topological prominences from a spectrogram via time-frequency bands. In this paper, we introduce the algorithm functioning with two experimental applications in a high-quality urban sound dataset and environmental audio recordings to describe how it works and how effective it is in handling the input data. Finally, we have posed two Python scripts that would reproduce the proposed case studies in order to ease the reproducibility of our audio fingerprinting system.
\end{abstract}

\section{Introduction}

In computer science, fingerprinting is a procedure that summarizes the input data by mapping it to a much shorter item \cite{Broder1993}. Similarly to human fingerprints, such transformation contains the essential information and properties of the original data, so it can be used to identify it among other samples \cite{Wagner1983}.

Regarding the acoustic field, audio fingerprinting is understood as an algorithm that extracts the main components taking into account the perceptual characteristics of the audio \cite{Cano2005concepts}. Most of the time, these techniques are applied over the spectrogram representation of the signal. Then, the pattern extraction is conducted by means of time domain \cite{Ramalingam2006}, frequency domain \cite{Seo2006}, or a combination of both called time-frequency domain \cite{Chun2002}. As far as implementation is concerned, there are some techniques created for this purpose, although they have their own advantages and limitations. The phase-based \cite{Arnold2014, Wang2015} and chroma-based \cite{Kim2008a, Kim2008b} fingerprinting techniques are widely used. In regard to data transformations, wavelets have been very effective in this field \cite{Baluja2008, Kamaladas2013, Jiang2019}. Nonetheless, this is not the only feature utilized for this purpose \cite{Burges2002, Seo2005, Miller2005, Serrano2022}.

For applicability purposes, \cite{wang2003} developed the idea of a constellation map for Shazam Entertainment in order to implement an audio search algorithm. Over the years, many different techniques have been developed \cite{Cano2005review}. However, it is worth mentioning that their implementation in machine learning tasks is very useful to reduce training costs, and so enable faster implementations. For example, we can also find recognition of activities of daily living via audio fingerprinting \cite{Pires2018}.

This paper presents the SpectroMap algorithm for creating audio fingerprints from a given audio signal. The method has been designed in order to deal with both raw audio excerpts and pre-processed spectrograms. The main objective is to cover the audio matching task because it can be considerably time-consuming \cite{Kurth2008, Garcia2019}. In essence, this paper is motivated by our previous work \cite{lopezgarcia2022}, where an in-depth example of the application of audio fingerprinting was applied to music plagiarism.

\section{Methodology}

\begin{figure}[ht!]
\centering
    \includegraphics[width=\linewidth]{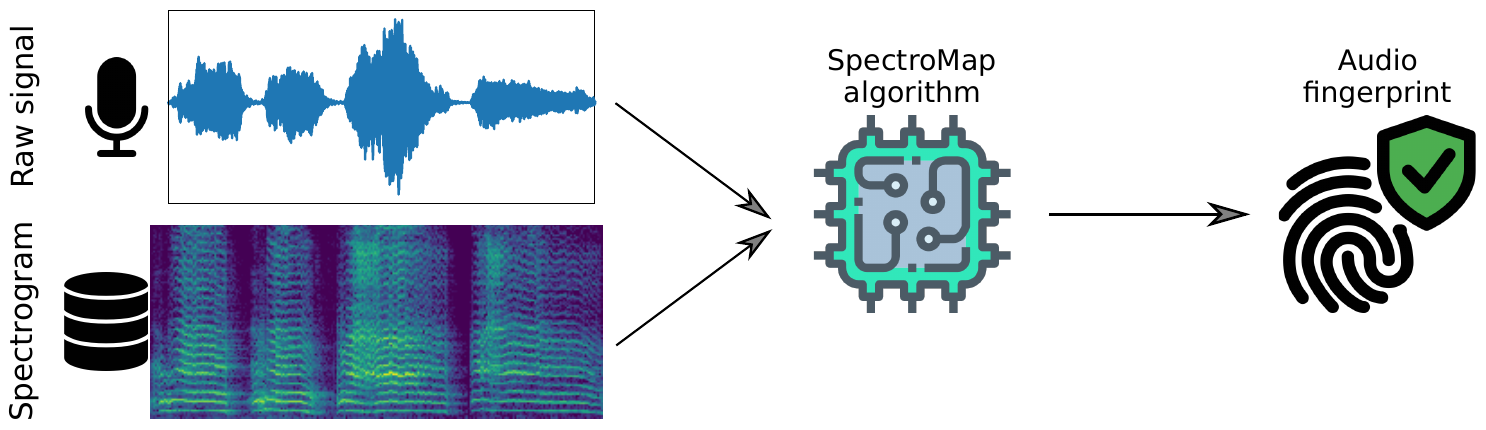}
    \caption{Architecture of the SpectroMap algorithm with the two possible input audio samples.}
    \label{fig:spectromap_architecture}
\end{figure}

The algorithm presented in this paper has been designed in order to carry out the entire process required to get the fingerprint from a given audio signal. In this manner, we facilitate open-source software able to produce large-scale signal processing. Depending on the objective of the user, we can use a raw signal or an already computed spectrogram as input when initializing the SpectroMap object. In case we decide to use raw signals, we can also include the required parameters for the signal processing step. Thereupon, the algorithm computes a local search to extract the topological prominences of the given spectrogram. The architecture of SpectroMap is depicted in Figure~\ref{fig:spectromap_architecture}. In this section, we have detailed the two steps that perform the fingerprint extraction of our algorithm.

\subsection{Signal processing}

With the aim of implementing a fingerprint extraction for a given musical signal $X_t$, we have designed an algorithm that computes a global peak detection over the spectrogram associated to give us its constellation map. Let $N_{FFT}$ and $N_O$ be the length of the Fast Fourier Transform (FFT) window and the number of elements to overlap between segments respectively, we first compute the spectrogram of the signal ($S_{tfa}$), by using the Hamming window, in order to get the (time, frequency, amplitude) vectors by considering these two parameters. Such representation contains the amplitude spatial information to analyze. Our engine search determines whether a time-frequency point can be considered locally relevant according to its neighborhood. Then, the detection is processed regarding a required band. Let $\{T_i\}_{i=1}^n$ and $\{F_j\}_{j=1}^m$ be the time and frequency bands of the spectrogram with the amplitude of the event, we can reformulate the spectrogram $S_{tfa} = (T_i)_{i=1}^n = (F_j)_{j=1}^m$ as its rows and columns representations.

As part of the engine search, we define two windows $\phi_T^{d_T}$ and $\phi_F^{d_F}$ to process the local pairwise comparisons with a respective length of $d_T$ and $d_F$, whose functionality is to extract a number of elements of the band and return the local maximum. Without limiting the generality of the foregoing, we can mathematically describe the time-band window mechanism with a length of $0< d_T\le n$ and structure $T_i = (T_i^1, ..., T_i^n)$ as:

\begin{equation}
    \displaystyle\phi_T^{d_T}(T_i) = \left(\max{\{T_i^k, \dots, T_i^{k+d_T}\}}\right)_{1\le k\le n-d_T-1},
    \label{eq:peak_search}
\end{equation}
per each band $i\in\{1,\dots,n\}.$

When we group all the values we drop those elements that have an equal index to avoid duplicates. Hence, we can group the window of each band to create the set:

\begin{equation}
    \displaystyle\Phi_T^{d_T} = \{\phi_T^{d_T}(T_i)\}_{i=1}^n.
    \label{eq:fingerprint}
\end{equation}

\begin{figure}[ht]
\centering
    \includegraphics[width=0.6\linewidth]{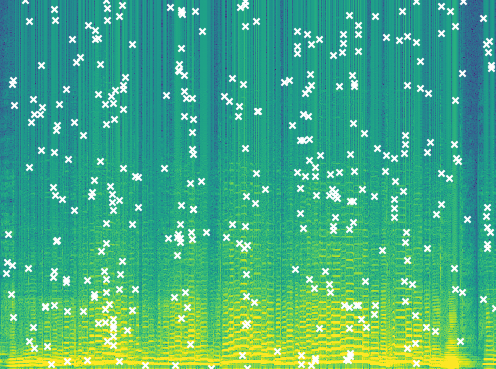}
\caption{Example of the spectrogram of an acoustic signal with its fingerprint stacked. The magnitudes are presented as seconds on the X-axis, Hertz on the Y-axis, and Decibels depicted as a color map.}
\label{fig:spectrogram}
\end{figure}

This way, we get the topologically prominent elements per each feature vector. Owing to the equation (\ref{eq:peak_search}), it is easy to note that even though there are $n-d_T-1$ matches, the window $\phi_T^{d_T}(T_i)$ may contain a smaller number of elements whenever $d_T > 2$. Depending on how restrictive we need to be, we can proceed with just one of the bands or combine them to create a more stringent search and distortion resistance since it is returned only the peaks that are prominent in both directions. Finally, the algorithm merges all the band-dependent peaks, as shown in equation (\ref{eq:fingerprint}), to give us the total number of spatial points that determine the so-called audio fingerprint.  In Figure~\ref{fig:spectrogram}, we can see a graphical example of an audio fingerprint.

\subsection{Algorithm}

\begin{figure}[t!]
\centering
    \includegraphics[width=\linewidth]{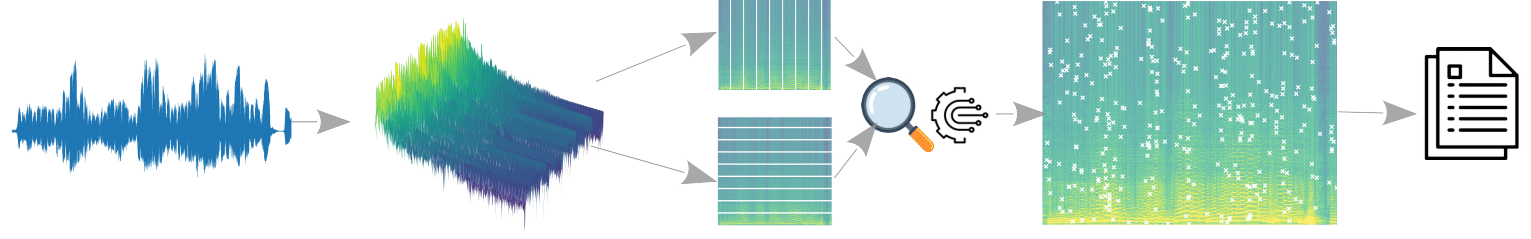}
\caption{Flowchart with the inner architecture of the algorithm implemented that detects the topological peaks of a spectrogram.}
\label{fig:algorithm}
\end{figure}

Our engine search, which boosts SpectroMap, processes audio signals in order to return an output file with the (time, frequency, amplitude) peaks detected in its spectrogram representation. Thus, it can be combined with the Mercury software to complete an in-depth comparison between music excerpts. Figure \ref{fig:algorithm} has a cursory description of the performance of SpectroMap. The algorithm basically batches the files by means of the following steps:

\begin{enumerate}[label={Step \arabic*}, align=left]
    \item Decide the window to use and set the parameters $N_{FFT}$ and $N_O$.
    \item Read the audio file to get its amplitude vector and its sample rate.
    \item Compute the spectrogram through the associated Fourier transformations.
    \item Set a fixed window length ($d_T$, $d_F$ or both) for the pairwise comparisons.
    \item\label{it:setting} Choose the settings to proceed with the peak detection over a selected band or a combination of both.
    \item Create an identification matrix consisting of a binary matrix with the same shape as the spectrogram with the position of the highlighted prominences.
    \item Extract such elements and create a file with the (time, frequency, amplitude) vectors.
\end{enumerate}

Regarding step~\ref{it:setting}, the authors highly recommend selecting both bands to perform the peak detection since the output is more filtered and spatially consistent. For the remainder steps, the choice is a personal decision that depends on the scope of the research. It is worth mentioning that the limitations of the method depend on the functionality of the Signal module of the SciPy library. Both installation and usage are described in our GitHub repository \cite{spectromap2022}.

\section{Case study: Processing of environmental and urban sound events}\label{sec:results}

The aim of this section is to present an experiment in which the performance of SpectroMap is analyzed in terms of computational cost. To this end, we have evaluated the speed of our algorithm over two datasets. On the one hand, Urban Sound 8K \cite{Salamon2014} is an audio dataset that contains 8732 labeled sound excerpts. The files are pre-sorted into ten folds in order to help in the reproduction and comparison of machine-learning experiments. The samples have a duration of $\approx4$s and they are classified as urban sounds from 10 classes: air conditioner, car horn, children playing, dog bark, drilling, engine idling, gunshot, jackhammer, siren, and street music. On the other hand, ESC-10 \cite{piczak2015} is a labeled collection of 400 environmental audio recordings suitable for benchmarking methods of environmental sound classification. In particular, ESC-10 is a subset of the major dataset ESC-50, which contains 2000 audio excerpts with a total size of $\approx600$MB publicly available (\url{https://github.com/karolpiczak/ESC-50\#download}).

For both datasets, the main use commonly attached is the classification task via supervised AI models. We can find robust performance ($94.6\%$ accuracy) utilizing CNN architectures \cite{Hassan2019} for the Urban Sound 8K set and other applications in low-cost monitoring devices \cite{Mydlarz2017}. For the ESC-50 dataset, and so the ESC-10, it has been shown that deep architectures such as Transformers \cite{Chen2022} and CNNs \cite{Elizalde2022} can learn with high precision ratios from this kind of audio sources with $97.00\%$ and $96.70\%$ of accuracy respectively.

\subsection{Computational costs}

Table~\ref{table:results} is presented the computational cost associated with the audio fingerprinting extraction task. All the experiments were produced by using seconds as time magnitude. For both datasets, it is computed the peak detection per folder and per audio sample. The Python script utilized to obtain the Table~\ref{table:results} is displayed in \ref{sec:python_implementation}. The computer that conducted the experiments was equipped with an AMD Ryzen 7 3700u with 16GB RAM running in Ubuntu 20.04.3 LTS OS.

\begin{table*}[t!]
\centering
\caption{Summary of the computational costs in seconds, with four significant figures, produced during the audio fingerprinting task for both Urban Sound 8K and ESC-10 experiments. The mean column contains the average plus-minus its standard deviation.}
\label{table:results}
\begin{tabular*}{\textwidth}{c @{\extracolsep{\fill}} crrrrrr}
\toprule
& & & \multicolumn{4}{c}{\textbf{Processing times (seconds)}} & \\
\cmidrule{4-7}
 & Set & Files &  Min &  \multicolumn{1}{c}{Mean} &  Max & Total &  it/s \\

\midrule
\multirow{10}{*}{\rotatebox[origin=c]{90}{\textbf{Urban Sound 8K}}}
&  1 &  874 & 0.0046 &  0.0885 $\pm$  0.0257 &  0.2414 &   77.4187 &  11.2892 \\
&  2 &  889 & 0.0047 &  0.0921 $\pm$  0.0382 &  0.2686 &   81.9609 &  10.8466 \\
&  3 &  926 & 0.0055 &  0.1154 $\pm$  0.0469 &  0.2682 &  106.9068 &   8.6617 \\
&  4 &  991 & 0.0098 &  0.0957 $\pm$  0.0385 &  0.2546 &   94.8629 &  10.4466 \\
&  5 &  937 & 0.0075 &  0.0905 $\pm$  0.0367 &  0.2606 &   84.8130 &  11.0478 \\
&  6 &  824 & 0.0079 &  0.1135 $\pm$  0.0486 &  0.2799 &   93.5494 &   8.8081 \\
&  7 &  839 & 0.0064 &  0.1040 $\pm$  0.0407 &  0.2265 &   87.3159 &   9.6087 \\
&  8 &  807 & 0.0094 &  0.1114 $\pm$  0.0502 &  0.3355 &   89.9648 &   8.9701 \\
&  9 &  817 & 0.0067 &  0.1277 $\pm$  0.0500 &  0.2494 &  104.3812 &   7.8270 \\
& 10 &  838 & 0.0087 &  0.0929 $\pm$  0.0350 &  0.2497 &   77.8634 &  10.7624 \\
\midrule
\multirow{5}{*}{{\rotatebox[origin=c]{90}{\textbf{ESC-10}}}}
& 1  &   80 & 0.1256 &  0.1803 $\pm$  0.0244 &  0.2859 &   14.4258 &   5.5456 \\
& 2  &   80 & 0.1434 &  0.1775 $\pm$  0.0127 &  0.2104 &   14.2069 &   5.6310 \\
& 3  &   80 & 0.1400 &  0.1799 $\pm$  0.0153 &  0.2335 &   14.3952 &   5.5573 \\
& 4  &   80 & 0.1169 &  0.1777 $\pm$  0.0193 &  0.2210 &   14.2231 &   5.6246 \\
& 5  &   80 & 0.1082 &  0.1781 $\pm$  0.0180 &  0.2142 &   14.2512 &   5.6135 \\

\bottomrule
\end{tabular*}
\end{table*}

\subsection{Graphical representation of the outputs for the environmental sound dataset}

\begin{figure}[b!]
\centering
    \includegraphics[width=\linewidth]{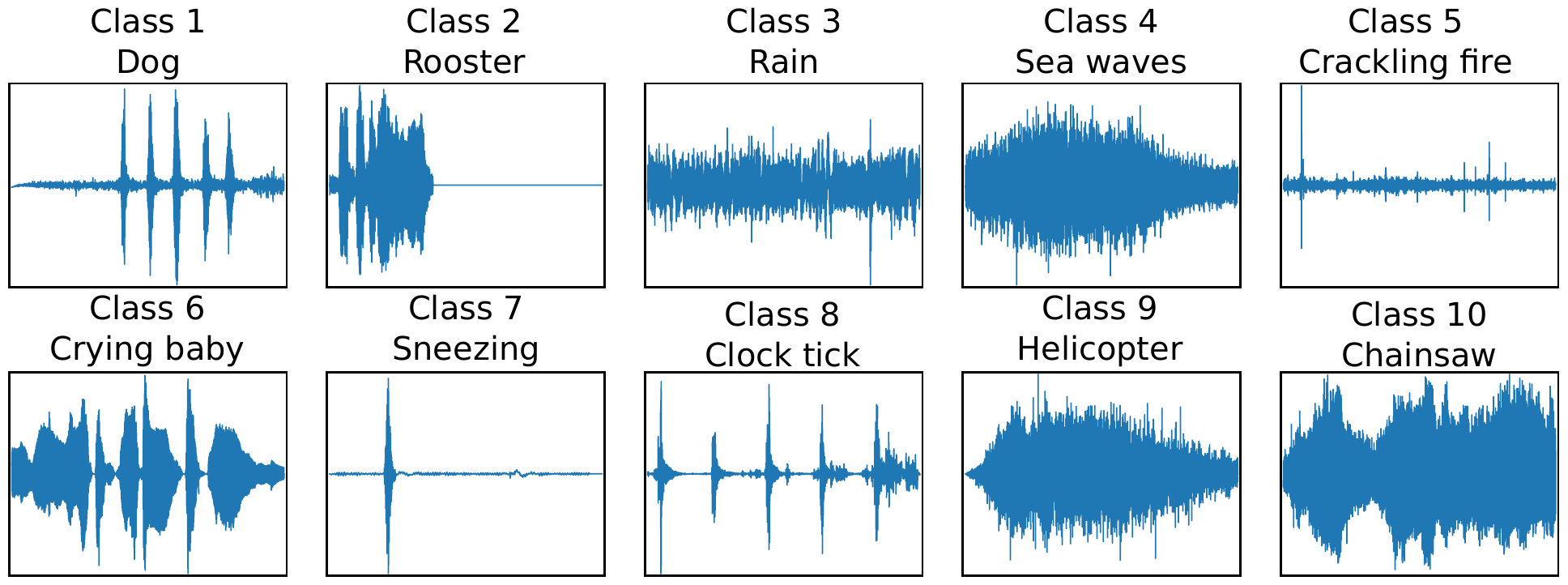}
\caption{Ten samples of the ESC-10 dataset representing each of the different acoustic classes.}
\label{fig:samples_ESC10}
\end{figure}

When we are conducting the signal processing stage for extracting the audio fingerprint of the audio samples, the representation of the fingerprints per class gives us significant information about the events. In order to give an overview of such a depiction, we have analyzed the classes of the ESC-10 dataset. The main point is to present the coordinates (time-frequency) relevant in terms of membership. On the one hand, figure~\ref{fig:samples_ESC10} contains a random sample per each of the 10 available classes. On the other hand, once we have carried out our algorithm, we have stored the coordinates that represent a peak within the fingerprint per each sample as a sequence $\{(t_i,f_i)\}_{i=1}^{I_n}$ so that each fingerprint contains $I_n$ topological prominences. With that information, we have generated a global class fingerprint consisting of natural entries that determine the number of times a coordinate has been selected as a peak per each sample of the same class. With the same notation as (\ref{eq:fingerprint}), we can define the global class fingerprint per each class $k$ as

\begin{equation}
    FP_k = \sum_{i=1}^{N_k}\mathbf{1}(\Phi_i^k),
\end{equation}

where the summation stands for the matrix sum operator and $\mathbf{1}$ for the matrix characteristic function of each fingerprint, $\Phi_i^k$ for the $i$ fingerprint of the class $k$, and $N_k$ for the number of elements in the class $k$.

\begin{figure}[t!]
\centering
    \includegraphics[width=\linewidth]{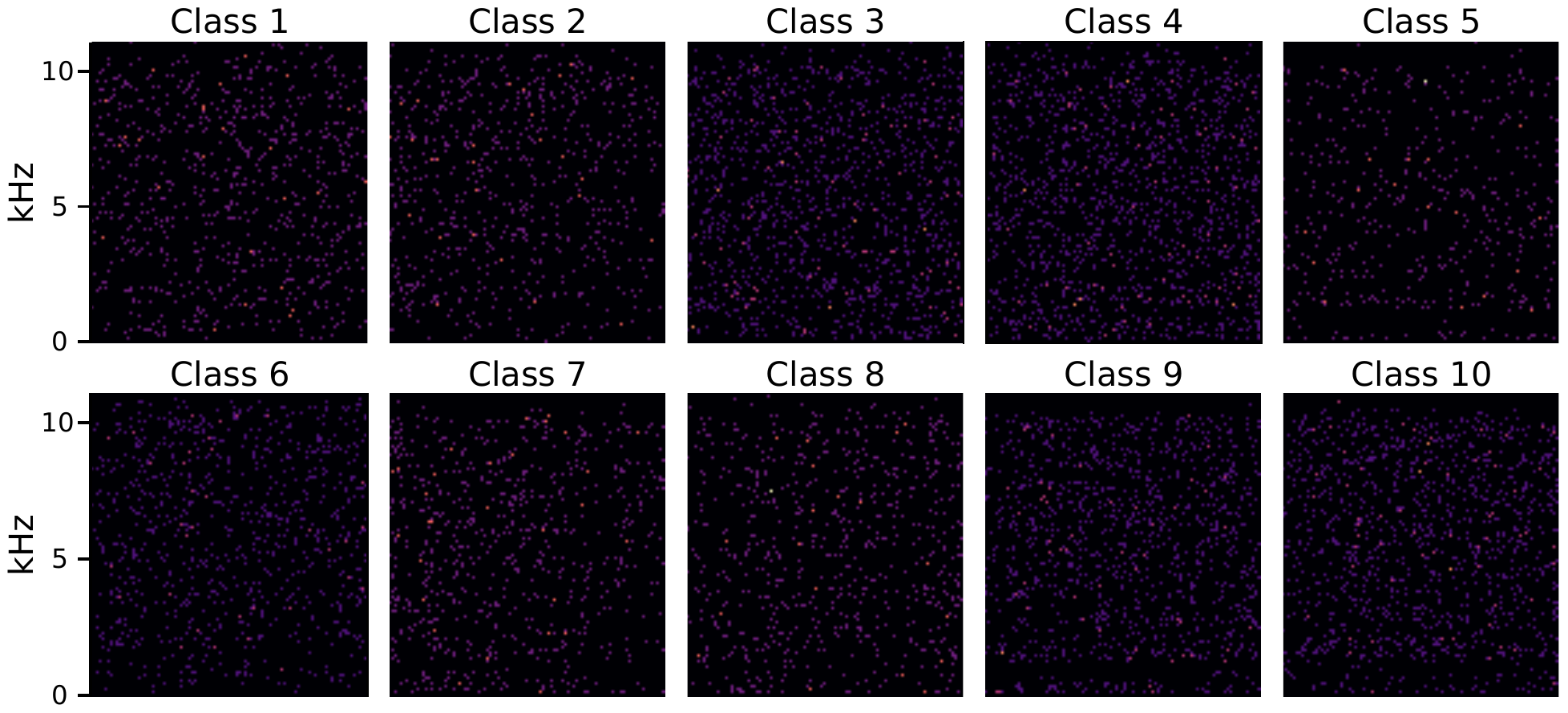}
\caption{Constellation map of the different classes of the ESC-10 datasets. Each subfigure contains the stacked aggregation per each fingerprint of each acoustic class indicating that the brighter the color, the more relevant the coordinate.}
\label{fig:fingerprints_ESC10}
\end{figure}

Considering all the mathematical notation aforementioned, Figure~\ref{fig:fingerprints_ESC10} shows each of the $FP_k$ in a viridis color palette indicating that brighter colors have a major impact in the representation of $FP_k$.

%
\section{Discussion and future work}

The SpectroMap algorithm has shown great performance when dealing with real-world acoustic scenarios. In the case studies conducted, our algorithm took a total of 899.03" (14' and 59.03") for the Urban Sound 8K and 71.50" (1' and 1.50") for the ESC-10 datasets. This can be summarized in a number of 9.82 $\pm$ 1.14 and 5.59 $\pm$ 0.03 iterations per second on average respectively. Therefore, SpectroMap can be considered an effective publicly available technique for audio fingerprinting.

One of the major advantages that arise from our experiments is that we can efficiently process audio signals (or even many others) for further analysis. Additionally, all the stages from which the acoustic sample is transformed are clearly defined, thus removing any kind of black boxes. Then, from these contributions, a potential future work would be to approach machine learning tasks by means of distance measures or similarity functions between audio samples. On the one hand, we could attach classification problems with a similar strategy to the KNN algorithm \cite{Evelyn1951, Cover1967, Jiang2007}. Basically, we would predict the class of some audio regarding its distance to the already known fingerprints. Another alternative would be the use of AutoEncoders \cite{Zhang2020}, with a semi-supervised approach, that reconstructs some audio \cite{Mnasri2020} from the information of a given set of fingerprints. On the other hand, we could perform an unsupervised strategy to determine the different sound sources based on the distribution that they present, such as K-means \cite{Lloyd1982} or DBSCAN \cite{Ester1996}.

Finally, it is important to remark that the use of Hertz as a frequency scale has been used for simplicity. Our main purpose has been to introduce the SpectroMap algorithm and show its applicability and performance. Then, we conducted a basic signal process to convert a signal into a spectrogram. However, there exist many choices to get different scales or units. For instance, the Mel-scale \cite{Stevens1937, oshaughnessy1987} would be a great alternative in order to get the perceptual scale of pitches of the events studied. A further application can be found in \cite{Shen2018} and \cite{Jang2019}).

%
\section{Conclusions}

We have introduced SpectroMap, a peak detection algorithm whose main application is the extraction of audio fingerprints. The algorithm not only processes raw signals but also preprocessed spectrograms, which means a major advantage in this field. Apart from a detailed explanation of the procedure and structure of the algorithm, we have also evaluated its performance in state-of-the-art datasets for audio analysis. It has been shown that SpectroMap is an effective and fast algorithm with an average of 1.340 and 3.336 iterations per second for the datasets presented in the case study (Urban Sound 8K and ESC-10). Further interpretations and representations have been shown in order to give a better understanding of the outputs of our algorithm. The code and Python implementation of the package has been presented in a straightforward manner in order to ease applicability and reproducibility. Even though we have not emphasized the underlying application on audio signals with comparison purposes, an instance of such an application can be found in our last paper \cite{lopezgarcia2022}.


\bibliographystyle{unsrt}

\begin{appendices}

\section{Python implementation}\label{sec:python_implementation}

This section is dedicated to the application of the SpectroMap algorithm to some kind of example. In particular, the module is designed to process either a raw signal or a spectrogram. For the first case, we make use of the \verb|spectromap| object. For the second case, we apply the \verb|peak_search| function. In addition, the script that reproduces the results shown is displayed at the end in order to easy reproducibility purposes.

The library was written with the Python 3.8 version and its usage depends just on NumPy 1.19 and SciPy 1.6.3. packages.
The repository is under the GNU General Public License v3.0.

\subsection{Application over a raw signal}

\begin{python}
import numpy as np
from spectromap.functions.spectromap import spectromap

# Generate a random signal
y = np.random.rand(44100)
kwargs = {'fs': 22050, 'nfft': 512, 'noverlap':64}

# Instantiate the SpectroMap object
SMap = spectromap(y, **kwargs)

# Get the spectrogram representation plus its time and frequency bands
f, t, S = SMap.get_spectrogram()

# Extract the topological prominent elements from the spectrogram.
# Coordinates matrix as (time, freq)
# Peak matrix.
fraction = 0.15 # Fraction of spectrogram to compute local comparisons
condition = 2   # Axis to analyze (0: Time, 1: Frequency, 2: Time+Frequency)
id_peaks, peaks = SMap.peak_matrix(fraction, condition)

# Get the peaks coordinates as as (s, Hz, dB)-array.
extraction_t_f_dB = SMap.from_peaks_to_array()
\end{python}

\subsection{Application over a given spectrogram}

\begin{python}
from spectromap.functions.spectromap import peak_search

fraction = 0.05 # Fraction of spectrogram to compute local comparisons
condition = 2   # Axis to analyze (0: Time, 1: Frequency, 2: Time+Frequency)
id_peaks, peaks = peak_search(spectrogram, fraction, condition)
\end{python}

\subsection{Application over a dataset}

Here is presented the script that reproduces the experimental cases for the Urban Sound 8K described in \autoref{sec:results}

\begin{python}
# OS packages
import os
import time
# Parallel computing
from pqdm.processes import pqdm

# Sinal processing modules
import numpy as np
import librosa
from spectromap.functions.spectromap import spectromap

NFFT = 1024
FRACTION = 1/3  # Fraction of spectrogram to compute local comparisons
CONDITION = 2   # Axis to analyze (0: Time, 1: Frequency, 2: Time+Frequency)

def spectromap_for_urbansound8K(file):
    # Transform file
    file = os.path.join(folder_path, file)
    # Read file
    signal, sample_rate = librosa.load(file, mono=True)
    kwargs = {'fs': sample_rate, 'nfft': NFFT}
    # Initialize SpectroMap
    t0 = time.time()
    SMap = spectromap(signal, **kwargs)
    # Get the spectrogram representation plus its time and frequency bands
    f, t, S = SMap.get_spectrogram()
    # Extract the topological prominent elements from the spectrogram
    id_peaks, peaks = SMap.peak_matrix(FRACTION, CONDITION)
    # Get the peaks coordinates as as (s, Hz, dB)-array.
    extraction_t_f_dB = SMap.from_peaks_to_array()
    cost_time = time.time() - t0
    return cost_time

# Path to Urban Sound 8K audio excerpts
path_UrbanSound8K = os.path.join('./', 'UrbanSound8K/audio')

computational_costs = []
FOLDERS = os.listdir(path_UrbanSound8K)
for folder in FOLDERS:
    print('\n\nProcessing: {}'.format(folder))
    folder_path = os.path.join(path_UrbanSound8K, folder)
    # Start process
    t0 = time.time()
    folder_time = pqdm(os.listdir(folder_path),
                       spectromap_for_urbansound8K,
                       n_jobs = os.cpu_count())
    folder_time = np.array(folder_time)
    # End process
    computational_costs.append(folder_time)

np.save('UrbanSound8K_computational_costs.npy', computational_costs)
\end{python}
\end{appendices}

\end{document}